# Response of Fractal Penetration of Magnetic Flux to Disorder Landscape in Superconducting Films

Zuxin Ye,[a] Qiang Li,[a,*] W. D. Si,[b] M. Suenaga,[a] V. F. Solovyov,[a] and P. D. Johnson[b]

[a]Materials Science Department, Brookhaven National Laboratory, Upton, New York 11973

[b]Physics Department, Brookhaven National Laboratory, Upton, NY 11973

Magnetic flux front and induction contours in superconducting $YBa_2Cu_3O_{7-\delta}$ films with defect size $s \sim \xi$ (superconducting coherence length) and $s \gg \xi$ are studied by magneto-optical imaging. Robust self-affine spatial correlation was observed using scaling analysis in the small pinning-dominated ($s \sim \xi$) disordered films. The roughness exponent $\alpha$ was determined to be ~ 0.66, independent of numbers of defects (or the film thickness). When the disorder landscape also included a distribution of large defects ($s \gg \xi$), the flux front and induction contours exhibited self-similarity, with a fractal dimension $D$ determined to be ~ 1.33 using the box-counting method. The remarkably different flux penetration patterns were shown to be the manifestation of self-organized criticality at different length scales.

[*] Author to whom correspondence should be addressed: electronic mail: qiangli@bnl.gov

## I. Introduction

The pattern formation in a physical system with time-independent quenched disorder and time-dependent thermal disorder has been a topic of great interest for many years.[1-3] Scaling laws are generally used to describe the dynamics of pattern formation in a wide range of these complex systems, including domain wall in ferromagnetic materials,[4] paper wetting[5] or burning,[6,7] deposition process,[2] flux propagation in type II superconductors.[8,9] It is particularly important to understand how a flux front negotiates the quenched disorder in a superconductor in the presence of driving forces and thermal fluctuations, both for elucidating the fundamental physics of vortex matter and the practical applications of superconductivity.[10,11] Flux penetration in type II superconductors was first suggested by C. Tang[12] to have the general features of self-organized criticality[13] (SOC), where, in response to an external perturbation, the flux pattern evolves between critical states via various sizes of avalanche-like relaxation processes, similar to a sand-pile. Magnetic flux penetration and relaxation are thus expected to follow certain spatial and temporal scaling laws, which are strongly influenced by the static quenched disorder of the material and thermal disorder at finite temperatures. However, numerous earlier experimental studies in both low and high $T_c$ superconductors did not produce consistent results through scaling analysis (see Ref. 9 and references therein). Recently, an interesting study of kinetic roughening of penetrating flux fronts in $YBa_2Cu_3O_{7-\delta}$ (YBCO) superconducting films was reported by Surdeanu *et al*.,[8] employing the dynamic scaling concepts used in the studies of interface roughening in stochastic systems. In their study, scaling analysis gave well defined self-affine (anisotropic self-similarity) exponents, consistent with the prediction of directed percolation depinning (DPD) model.[2] Furthermore, Vlasko-Vlasov *et al*. recently showed in niobium films that not only the flux front but also the magnetic induction profiles carry the fingerprints of SOC.[9] Vortex matter in a superconductor is in fact an excellent model system for studying the general nonlinear diffusion process, because a tailored defect structure and defect landscape can be synthesized in superconductors, allowing us to investigate the nonlinear diffusion process at different length scale.

In the studies of flux pattern formation in superconductors, mostly considered disorders are the randomly distributed small defects, i.e. size $s \sim$ superconducting coherence length $\xi$, acting as effective vortex pinning centers. In this scenario, the scaling laws directly reflect the competition between the vortex-vortex interaction and the vortex pinning by these defects. The former tends to maintain the long-range correlation of vortex lattice and to straighten the flux front, while the latter tends to hold the vortices in the defect locations and to roughen the flux front. A very different situation may exist when the disorder is much greater than $\xi$. In the case of $s \gg \xi$, a disordered region can no longer impede vortices locally. Such a characteristic alteration in the disorder landscape is expected to change the flux patterns dramatically.[14]

The purpose of the present study is to investigate, using magneto-optical imaging (MOI) technique, the flux pattern formation at quite different length scale in the superconducting films with various quenched disorders, not only those pinning defects but also large size defects. In the pinning disorder dominated films, we found that the spatial correlation function of the flux front and the magnetic induction profiles show excellent power law behavior. The obtained roughness exponent $\alpha \sim 0.66$ is consistent with the previous studies reported by various groups.[8, 9] Furthermore, the self-affine fractals were found to be independent of the film thickness $d$. However, in the films containing large defects, the flux pattern exhibited clear self-similarity. Applying the box-counting method, the fractal dimension $D$ was determined to be ~ 1.33 for both the flux front and the magnetic induction contours.

## II. Experimental Details

Epitaxial YBCO films were grown on single crystalline $LaAlO_3$ substrates using pulsed laser deposition (PLD), which is known to produce the films with small regions of disorder (for instance, dislocations), mostly of a flux pinning type.[15, 16] The *c*-axis oriented YBCO films with large-size defects were grown on single crystalline $SrTiO_3$ substrates using the $BaF_2$ *ex-situ* process, which produces large precipitates. Insulating $Y_2O_3$ or CuO particles have a physical size $s$ range 0.1 μm $< s <$ 3 μm spaced at ~ a few μm apart and randomly dispersed in the YBCO matrix.[17] These large precipitates do not pin the vortices. However, the matrix of the $BaF_2$ processed films is believed to be similar to the pinning dominated PLD films. This is because films processed by PLD and $BaF_2$ process have superconducting transition temperature $T_c$ above 91 K ($\Delta T_c \leq 0.3$ K). In addition, similar high critical current densities $J_c$ (~ $1.5 \times 10^{11}$ A/cm$^2$ at 4.2 K and self field) and similar $J_c(B)$ behavior for the 1 μm-thick films made by both methods suggest a comparable flux pinning strength.

Two PLD-processed YBCO films ($d$ = 0.2 and 1 μm) were patterned into 0.8 and 0.4 mm wide strips (each ~ 4 mm long), respectively. Knowing that flux penetration usually displays a heavy branching pattern in $BaF_2$-processed YBCO films,[18] we patterned the 1 µm-thick BaF$_2$-processed YBCO film into a circular disk of 5.3 mm diameter, leaving enough room for the fractal structure to develop during the flux propagation. Flux penetration patterns were directly captured using magneto-optical imaging (MOI) technique.[19] A detailed description of the low temperature MOI station used in this study can be found in Ref. 20. In brief, a high-resolution MOI indicator film was placed directly onto the sample surface, with magnetic field always applied perpendicular to the film surface, and polarizer and analyzer crossed at 90 degrees. Brightness intensity in a MO image represents the local magnetic field induction component normal to the surface of the film.

## III. Results and Discussions

**a) Self-affine fractals, small size disorders, and thickness of the films**

Fig. 1a and 1c are two typical MO images of the flux penetration pattern in the zero-field-cooled (ZFC) 0.2 µm-thick (a) and 1 µm-thick (c) PLD films taken at $T = 20$ K under an external field $B_a = 0.1$ T. As expected from the critical state models, the flux density gradually decreases from the edge to the flux front. The flux front, the border between the flux-free and flux-occupied region, can be extracted from MO images.[21] Fig. 1b and 1d show the lateral wandering of the flux fronts extracted from Fig. 1a and 1c, respectively. The flux fronts were plotted as a function $h(x)$ of the horizontal coordinate $x$ along the long edges of the strips, and the mean value of $h(x)$ was chosen as zero point on the $h$ axis. The self-affinity in the spatial correlation of these flux fronts was examined by analyzing the two-point correlation function $C(l)$.

The morphology of a self-affine interface, which resembles itself under anisotropic spatial transformation, follows the Family-Vicsek scaling relations.[1-3] Here we consider the spatial correlation of a self-affine interface embedded in a two-dimensional Euclid space, which represents a flux front wandering in the film. $C(l)$ is defined as

$$C(l) \equiv \{<[h(x+l) - h(x)]^2>_x\}^{1/2} \quad (1)$$

where $l$ is the separation of two points on the self-affine interface, and the average $<...>_x$ is taken over the whole observation window. The roughness exponent $\alpha$ can be determined from the following scaling relation

$$C(l) \sim l^\alpha \quad (2)$$

with $l$ being the variable. The roughness exponent $\alpha$, combined with other spatial or temporal exponents, characterizes the universality class of the interface evolution.

Fig. 2a shows the correlation function $C(l)$ of the flux fronts in the 0.2 μm-thick (open symbols) and 1.0 μm-thick (solid symbols) PLD-processed films at $T = 20$ K, and $B_a = 0.1$ T. A clear power law behavior was observed at $l < 30$ μm. $\alpha$ was found to be 0.67 and 0.66 for the 0.2 μm thick and 1.0 μm thick films, respectively. Within the framework of self-affine fractals, a fractal dimension $D = 2 - \alpha$ can be derived from the measured $\alpha$ value: $D \sim 1.33$. The derived $\alpha$ value in these PLD films is consistent with the predicted values in the DPD model. It needs to be emphasized that the DPD model is in the universality class of the quenched Kardar-Parisi-Zhang (KPZ) model. In the original model, the KPZ equation describes the motion of a driven interface subject to temporal disorder alone, and predicts a roughness exponent $\alpha = 0.5$.[22] However, in the quenched KPZ model (QKPZ), the temporal noise term is replaced by a static noise term, which predicts an $\alpha$ value ~ 2/3.[2] These predictions have been shown to be in a good agreement with earlier studies on YBCO thin films[8] and niobium thin films.[9] It is interesting to note that although the flux front in the thicker film (1 µm-thick) is considerably rougher (larger $C(l)$ values) due to a higher number of pinning centers for each vortex, the scaling law works quite well there with an $\alpha$ similar to the value in the thin film at virtually the same measuring length scale $l$.

**B. Temperature and field dependence of self-affine fractals**

The robustness of the scaling laws in the pinning dominated films was further explored by varying $T$ and $B_a$. The DPD-like roughness exponents observed in this work are the consequence of the flux propagation in a random medium dominated by quenched disorder. Since the thermal fluctuation of vortex lines leads to a dynamical sampling and hence averaging of the disorder potential over the spatial extent of the thermal displacement of the vortices,[10] the scaling laws are expected to change as $T$ increases. The rise in $T$ will result in a smoother flux front and a lower roughness exponent $\alpha$. Fig. 2b and 2c show the roughness exponent as a function of the normalized applied field $B_a/B_d$ at various $T$ for the PLD films, where $B_d$ is the characteristic field, given by $B_d = \mu_0 J_c d/2$. For a superconducting thin film under a perpendicular magnetic field, the $B_a/B_d$ ratio varies with the flux front mean position.[11, 19] A flux front with a small $B_a/B_d$ is close to the film edge, and hence is strongly influenced by the microscopic geometry of the film edge. For a photo-lithographically patterned film, the film edge tends to be very straight, which results in a small $\alpha$ value ($\alpha$ ~ 0 at edge) for the flux front near the edge. As $B_a/B_d$ increases, the flux front moves inside, and the edge geometry has less effect on the morphology of the flux front, with $\alpha$ eventually reaching a plateau where the $\alpha$ value corresponds to the intrinsic roughness exponent. As shown in Fig. 2b, for the 0.2 μm-thick film, the $\alpha$ value at the plateau is nearly constant for $T \leq 40$ K. At $T = 60$ K, the $\alpha$ value was significantly reduced to ~ 0.47. This reduction of $\alpha$ at elevated $T$ may suggest the crossover from the quenched disorder dominated region to the thermal fluctuation dominated region. In contrast, the $\alpha$ value at the plateau for the 1 μm-thick film remains nearly constant at $T$ up to 60 K. In the 1 μm-thick film, with a higher number of pins per vortex, thermal fluctuation appears to have less effect than in the 0.2 μm-thick film.

**C. Self-similar fractals and large-size disorders**

Fig. 3 shows a MO image of flux penetration in the ZFC BaF$_2$-processed YBCO film taken at $T = 4.2$ K and $B_a = 0.1$ T. (Note: only a ~ 60° portion of the disk was shown in the image due to the limited size of the MO indicator film). The flux pattern exhibits self-similar fractal with plenty of branching and overhang structures. A self-similar fractal pattern resembles itself under isotropic spatial transformations. Hence, the flux front is no longer single-valued. This can be easily understood by drawing a straight line from the disk center to the edge, where this line intersects the flux front at multiple points, i.e. the flux front has overhangs. Therefore, $C(l)$ is not suitable for the scaling analysis of a self-similar fractal pattern. Instead, we applied the box-counting method[1-3] to the flux front, as well as to a series of induction contour lines behind it with a fixed magnetic induction value. The procedure of box counting is as follows. A figure containing the flux front or a contour line extracted from MO images is broken into small boxes with size $l$. The number of boxes containing a part of the self-similar flux front (or contour) $N(l)$ is expected to follow a scaling law given as the following,

$$D \equiv \lim[\ln N/\ln(1/l)]_{l \to 0} \quad (4)$$

where $D$ is the fractal dimension.

Fig. 4a shows box-counting result of the flux front extracted from Fig. 3, where an excellent power law was clearly demonstrated. From the slope of the $N(l)$ curve, the fractal dimension was determined to be 1.33. Furthermore, the induction contours behind the flux front were found to follow similar scaling behavior. Using the same box counting method, the fractal dimension $D$ for all the flux contours was found to be close to 1.33. The inset to Fig. 4a shows the values of $D$ for the contours at various induction values $B_c$ for $B_a = 0.1$ T. To investigate the influence of thermal fluctuation on the flux pattern, we also performed the box-counting analysis on the flux fronts taken at $4.2 \leq T \leq 70$ K. In Fig. 4b, we plotted the fractal dimension $D$ as a function of the normalized applied field $B_a/B_d$. Except some reduction due to the edge effect at low field, $D$ shows a nearly constant value around 1.33 for all $T$. This $T$-independent $D$ is likely due to the fact that the typical spatial extent of thermal fluctuation is much smaller than the defect size in this film, and hence is unable to change the overall quenched disorder landscape and flux pattern significantly. The self-similar fractals with a constant fractal dimension $D$ observed in both the flux front and the induction contours demonstrate the robustness of such scaling behavior in the porous films. By closely comparing the pattern formation of flux penetration in the $BaF_2$ processed superconducting films with other known systems, we notice similar features in the percolating fluid patterns in porous medium, as well as domain walls propagation pattern. The common behavior among them suggests that the flux penetration in the $BaF_2$ processed films may be mapped onto a site percolation, where the local flux penetration is determined only by the local flux pinning strength in the critical state. This is similar to the invasion percolation of fluid flow in porous media[23] and random field Ising model of domain wall propagation in ferromagnetic materials.[24]

**D. Response of self-affine fractals and self-similar fractals to disorder size**

To explain the different fractal behaviors and associated scaling laws in the flux pattern formation in pinning dominated and $BaF_2$ processed YBCO films, we need to consider the relationship between the observed scaling laws and the two relevant length scales, the measurement length $l$ and the quenched disorder size $s$. $l$ is the spatial variable used in the scaling analysis such as $C(l)$ analysis and box-counting. At very small $l$, one sees more of the flux pattern in details, less of the correlation. The upper limit of $l$ is the observation window size $L$ (~ 6 mm in our experiment), while the lower limit is the spatial resolution of MOI (~ 1 μm). $s$ is the quenched disorder size determined by the microstructure of the superconductor. In the quenched disorder dominated films, the scaling law gives the DPD-like roughness exponent corresponding to the self-affine fractal for $l \gg s$. This is what we observed in our PLD films, and is consistent with previous studies in Ref. 8 and 9. To approach the regime of $l \sim s$, one has to either reduce the measuring length $l$, or increase the defect size $s$. For the PLD films, $l$ has to be reduced to the size of the disorder ~ $\xi$ for YBCO (~ 1 nm), clearly beyond the capability of current MOI technique. In addition, the magnetic core size of vortex in YBCO (measured by the penetration depth $\lambda$, ~ a few 100 nm) averages out the details at the disorder length scale. This perhaps is the reason why self-similar fractals have never been reported of the flux patterns in superconductors with only small pinning defects. On the other hand, by increasing the defect size, as in the case of the $BaF_2$ processed YBCO films, we were able to approach the regime of $l \sim s$ where a self-similar spatial correlation in flux pattern is observed with $D \sim 1.33$. This fractal dimension is virtually the same as that found in PLD films showing the DPD-like self-affine flux fronts. Therefore, it appears that the different fractal patterns revealed in the two cases are in fact governed by the same SOC behaviors, but manifest at different observation length scale, i.e. self-similarity at small $l$ regime ($l \sim s$), and self-affinity at large $l$ regime ($l \gg s$).

## IV. Conclusions

In conclusions, we observed the strikingly different fractal patterns of magnetic flux in superconducting films corresponding to the different landscape of random quenched disorders. For very small defects, the self-affine flux front and induction contours followed the robust scaling law given by the quenched KPZ model. For large defects close to the observation length, the flux penetration shows self-similar fractal patterns. Though the scaling law varies with length scale, the fractal dimensions are the same. This suggests that the different fractal behavior manifests the multiple aspects of the self-organized criticality at different length scales in flux pattern formation. Through tuning the relative disorder length scale, flux pattern formation offers a new way to explore the nature of the self-organized criticality in quenched disordered systems.

ACKNOWLEDGMENT

This work was supported by the U. S. Dept. of Energy, Office of Basic Energy Science, under contract No. DE-AC-02-98CH10886.

**FIGURE CAPTIONS**

Fig.1 MO images of flux penetration patterns and the lateral wandering of flux fronts in two ZFC YBCO films (edge indicated by arrows) grown by PLD. (a) and (b) MO image of the 0.2 μm-thick film driven by a perpendicular magnetic field of 0.1 T at $T = 20$ K, and corresponding lateral wandering $h(x)$ of the flux front. (c) and (d) MO image of the 1μm-thick film at the same field and temperature, and the corresponding $h(x)$.

Fig. 2 (a) The correlation function $C(l)$ of the flux fronts in the 0.2 and 1 μm-thick PLD processed YBCO films at $T = 20$ K and $B_a = 0.1$ T showing a power law behavior with the roughness exponent $\alpha = 0.67$ and 0.66, respectively. (b) and (c) The field dependence of $\alpha$ at various temperatures in (b) the 0.2 μm-thick, and (c) the 1μm-thick films.

Fig. 3 MO images of flux penetration patterns in a 1μm-thick YBCO film grown by $BaF_2$ process taken at $T = 4.2$ K under an applied field $B_a = 0.1$ T. The sample was patterned into a circular disk 5.3 mm in diameter.

Fig. 4 (a) The number $N$ of boxes needed to cover a flux front in the 1μm-thick $BaF_2$ processed YBCO film ($T = 4.2$ K and $B_a = 0.1$ T) as a function of box size $l$ in the box-counting method, showing a power law with the fractal dimension $D = 1.33$. Inset: the fractal dimension $D$ as a function of the magnetic induction contour value $B_c$. (b) Fractal dimension as a function of normalized applied field $B_a/B_d$ at various temperatures, where $B_d$ is the characteristic field proportional to the critical current density.

Fig.1

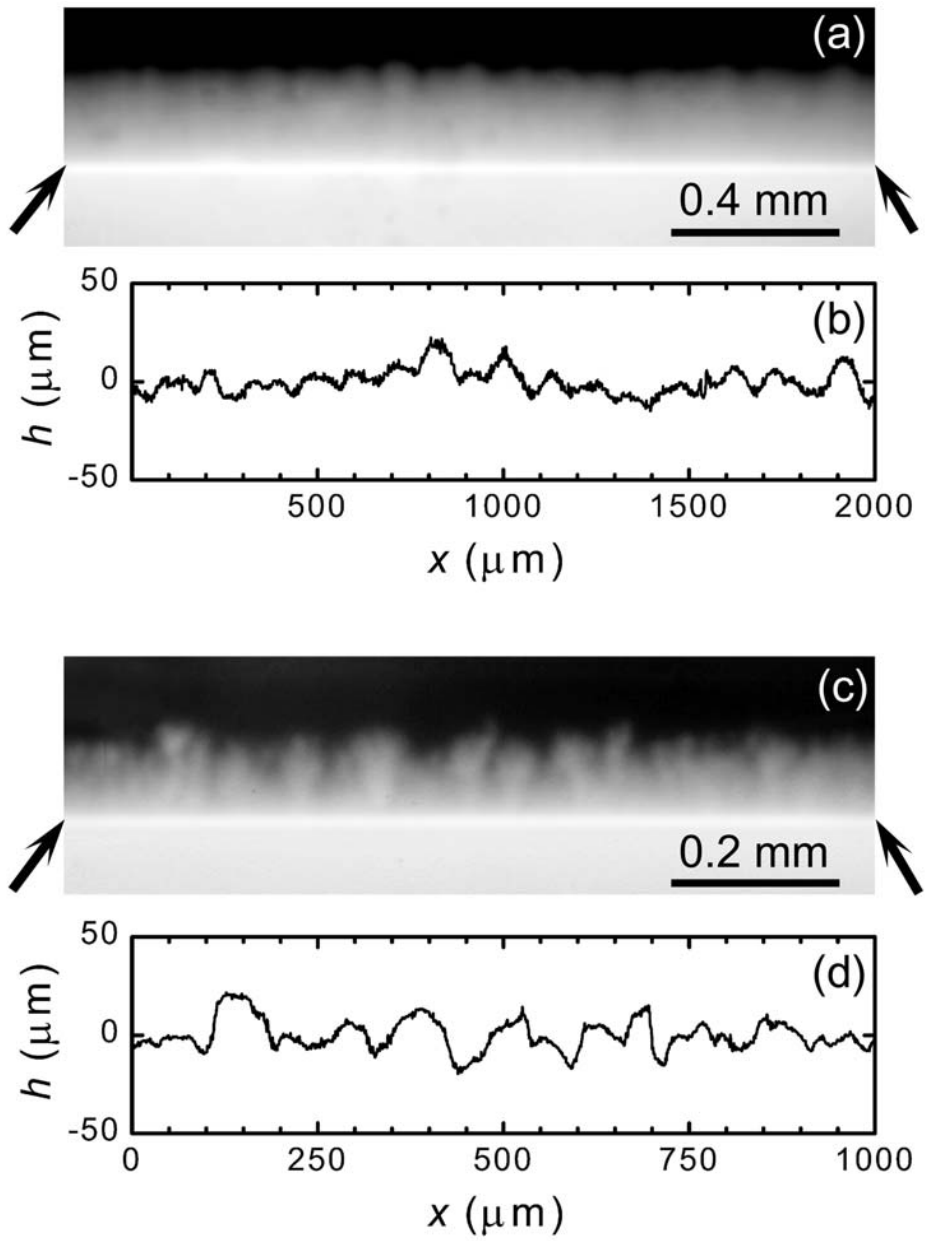

Fig. 2

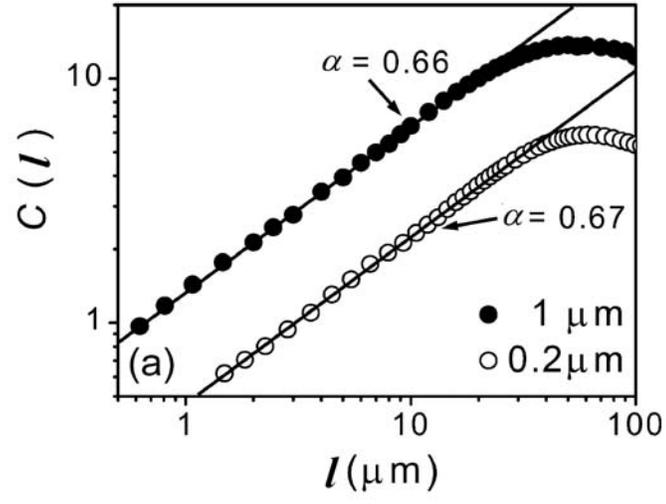

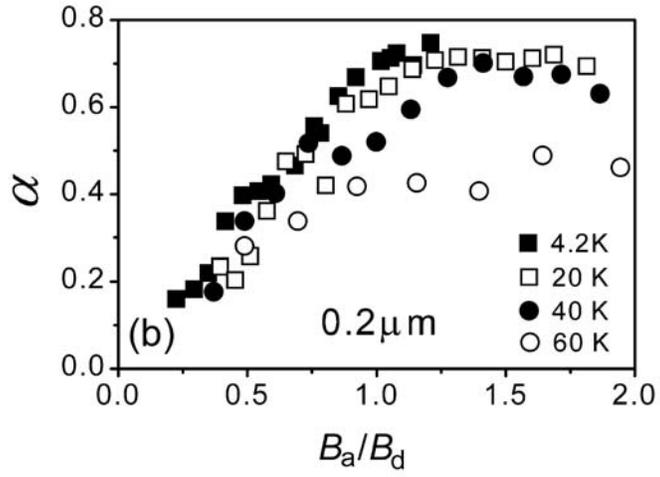

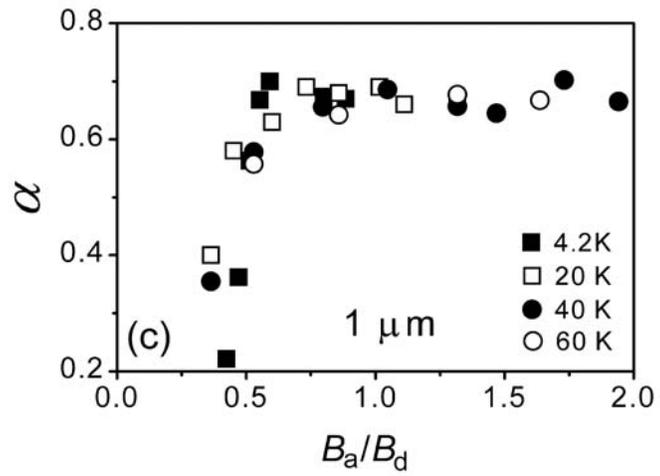

Fig. 3

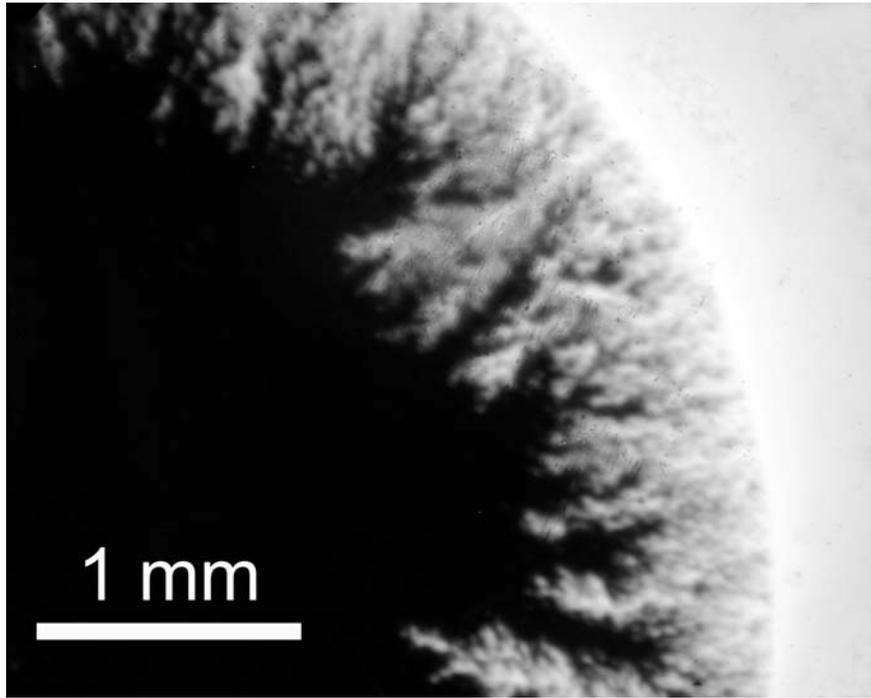

Fig. 4

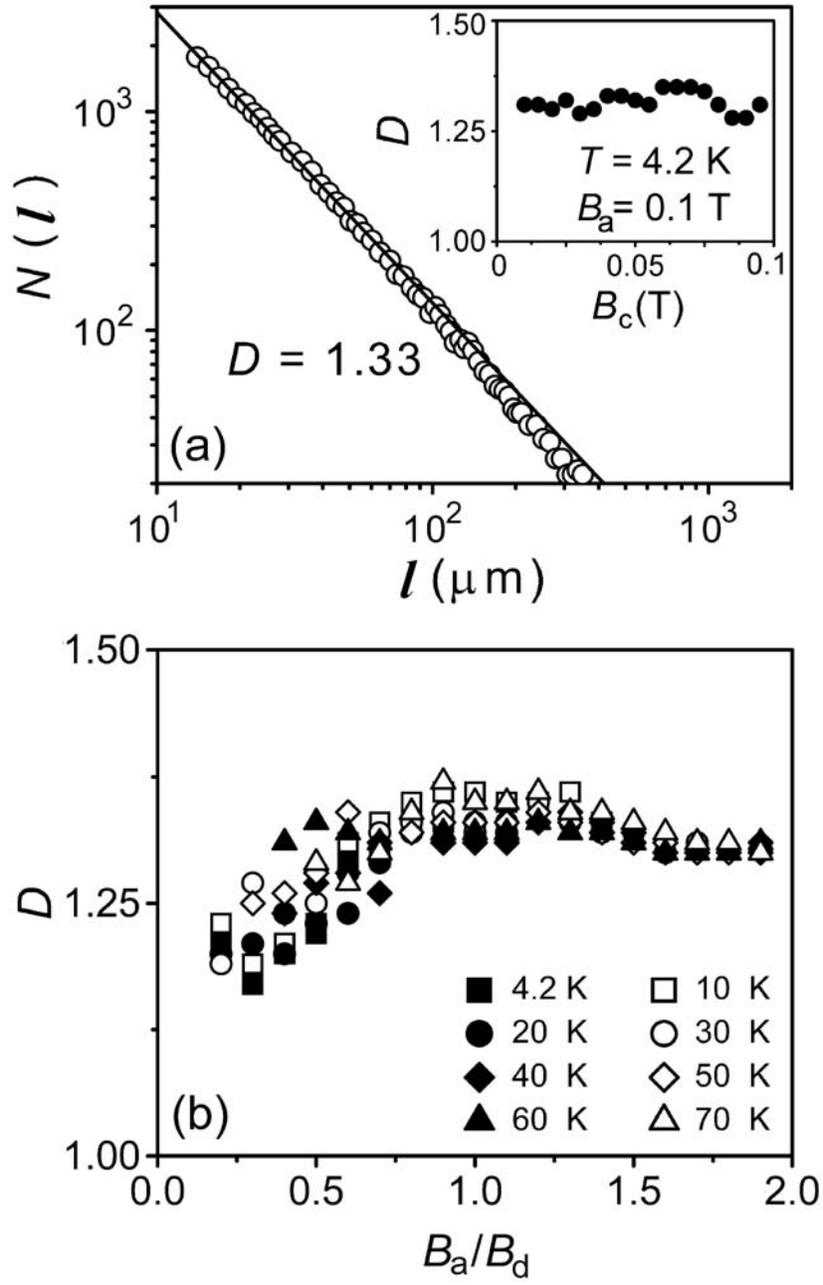